\documentclass[]{aa}
\usepackage{graphics}
\usepackage{txfonts}
\usepackage{amssymb}
\usepackage{lscape}
\usepackage{epsfig}
\usepackage{natbib}

\newcommand{\mincir}{\raise
-2.truept\hbox{\rlap{\hbox{$\sim$}}\raise5.truept\hbox{$<$}\ }}
\newcommand{\magcir}{\raise
-2.truept\hbox{\rlap{\hbox{$\sim$}}\raise5.truept\hbox{$>$}\ }}

\begin{document}

\title{The XXL Survey\thanks{Based on observations obtained with XMM-Newton, an ESA science 
mission with instruments and contributions directly funded by
ESA Member States and NASA.}} 

\subtitle{XXXV. The role of cluster mass in AGN activity}
\author{E. Koulouridis\inst{1} \and M. Ricci\inst{2} \and P. Giles\inst{3} \and C. Adami\inst{4} \and M. Ramos-Ceja\inst{5}  \and M. Pierre\inst{1} \and M. Plionis\inst{6,7} \and C. Lidman\inst{8} \and I. Georgantopoulos\inst{9} \and L. Chiappetti\inst{10} \and A. Elyiv\inst{11,12} \and S. Ettori\inst{13,14} \and L. Faccioli\inst{1} \and S. Fotopoulou\inst{15} \and F. Gastaldello\inst{10}  \and F. Pacaud\inst{5} \and S. Paltani\inst{16}  \and C. Vignali\inst{12}}

\institute{AIM, CEA, CNRS, Universit\'e Paris-Saclay, Universit\'e Paris Diderot, Sorbonne Paris Cit\'e, F-91191 Gif-sur-Yvette, France
\and Laboratoire Lagrange, Universit\'e C\^ote d’Azur, Observatoire de la C\^ote d’Azur, CNRS, Blvd de l'Observatoire, CS 34229, 06304, Nice cedex 4, France
\and School of Physics, HH Wills Physics Laboratory, Tyndall Avenue, Bristol, BS8 1TL, UK
\and LAM, OAMP, Universit\'e Aix-Marseille, CNRS, P\^ole de l'\'Etoile, Site de Ch\~{a}teau Gombert, 38 rue Fr\'ed\'eric Joliot-Curie, 13388, Marseille 13 Cedex, France
\and Argelander-Institut f\"ur Astronomie, University of Bonn, Auf dem H\"ugel 71, 53121 Bonn, Germany
\and National Observatory of Athens, Lofos Nymfon, GR-11810 Athens, Greece
\and Physics Department of Aristotle University of Thessaloniki, GR-54124, Thessaloniki, Greece
\and Australian  Astronomical  Observatory,  North  Ryde,  NSW  2113, Australia
\and Institute for Astronomy \& Astrophysics, Space Applications \& Remote Sensing, National Observatory of Athens, GR-15236 Palaia Penteli, Greece
\and INAF - IASF - Milano, Via Bassini 15, I-20133 Milano, Italy
\and Main Astronomical Observatory, Academy of Sciences of Ukraine, 27 Akademika Zabolotnoho St., 03680 Kyiv, Ukraine
\and Dipartimento di Fisica e Astronomia, Universit\'a di Bologna, via Gobetti 93/2, I-40129 Bologna, Italy
\and INAF - Osservatorio  Astronomico  di  Bologna,  via  Pietro  Gobetti 93/3, I-40129 Bologna, Italy
\and INFN, Sezione di Bologna, viale Berti Pichat 6/2, I-40127 Bologna, Italy
\and Center for Extragalactic Astronomy, Department of Physics, Durham University, South Road, Durham, DH1 3LE, United Kingdom
\and Department of Astronomy, University of Geneva, ch. d'Ecogia 16, CH-1290 Versoix, Switzerland}
\date{Received/Accepted}

\abstract{We present the results of a study of the AGN density in a homogeneous and well-studied sample of 167 bona fide X-ray galaxy clusters ($0.1<z<0.5$) from the XXL Survey, from the cluster core to the outskirts (up to $6r_{500}$). The results can provide evidence of the physical mechanisms that drive AGN and galaxy evolution within clusters, testing the efficiency of ram pressure gas stripping and galaxy merging in dense environments.}{The XXL cluster sample mostly comprises poor and moderately rich structures ($M=10^{13}$ -- $4\times10^{14} M_{\sun}$), a poorly studied population that bridges the gap between optically selected groups and massive X-ray selected clusters. Our aim is to statistically study the demographics of cluster AGNs as a function of cluster mass and host galaxy position.}
{To investigate the effect of the environment on  AGN activity, we computed the fraction of spectroscopically confirmed X-ray AGNs ($L_{\rm X [0.5-10\,keV]}>10^{42}$ erg cm$^{-1}$) in bright cluster galaxies with $M_i^*-2<M<M_i^*+1$, up to $6r_{500}$ radius. The corresponding field fraction was computed from 200 mock cluster catalogues with reshuffled positions within the XXL fields. To study the mass dependence and the evolution of the AGN population, we further divided the sample into low- and high-mass clusters (below and above $10^{14} M_{\sun}$, respectively) and two redshift bins (0.1--0.28 and 0.28--0.5).}
{We detect a significant excess of X-ray AGNs, at the 95\% confidence level, in low-mass clusters between $0.5r_{500}$ and 2$r_{500}$, which drops to the field value within the cluster cores ($r<0.5r_{500}$). In contrast, high-mass clusters present a decreasing AGN fraction towards the cluster centres, in agreement with previous studies. The high AGN fraction in the outskirts is caused by low-luminosity AGNs, up to $L_{\rm X [0.5-10\,keV]}=10^{43}$ erg cm$^{-1}$. It can be explained by a higher galaxy merging rate in low-mass clusters, where velocity dispersions are not high enough to prevent galaxy interactions and merging. Ram pressure stripping is possible in the cores of all our clusters, but probably stronger in deeper gravitational potentials. Compared with previous studies of massive or high-redshift clusters, we conclude that the AGN fraction in cluster galaxies anti-correlates strongly with cluster mass. The AGN fraction also increases with redshift, but at the same rate with the respective fraction in field galaxies.}{}

\authorrunning{E. Koulouridis et al.}
\titlerunning{The XXL survey - XXXV}

\maketitle

\section{Introduction}

Since the discovery that all massive galaxies in the local Universe host a central super massive black hole (SMBH) with a mass proportional to that of the galaxy spheroid (known as
the Magorrian relation) \citep{Magorrian98,Gultekin09,Zubovas12}, the study of SMBHs and active galactic nuclei (AGNs) has been a lively topic in modern astrophysics. To explain this interactive co-evolution, theoretical models proposed an AGN-driven feedback wind as a mechanism to regulate the amount of gas in galaxies \citep{Schawinski06,Cen11}.
Therefore, an accurate census of  AGNs is essential in understanding the cosmic history of accretion onto SMBHs and their relation to the host galaxy.

However, there is still no consensus about the mechanisms that trigger or suppress  AGNs. There is evidence that galaxy mergers and interactions play an important role in triggering AGNs \citep{Koulouridis06,Hopkins08}, but on the other hand, AGNs are also found isolated and undisturbed. Nevertheless, there is compelling evidence that the presence of AGNs is closely linked to the large-scale environment. As structures grow hierarchically, galaxies are accreted by progressively more massive dark matter halos, and  the majority of galaxies end up in groups and clusters \citep{Eke04,Calvi11}, which are therefore the predominant environment of galaxies and can play a very important role in establishing galaxy properties. As the most massive self-gravitating entities of the universe, clusters are also ideal laboratories for  investigating the impact of dense environments on AGN demographics. 

Clusters and groups are usually identified by optical and infrared surveys 
as concentrations of red-sequence galaxies \citep[e.g.][]{Gladders00,Koester07,Hao10,Rykoff14,Bleem15} or galaxy overdensities in photometric redshift space 
\citep[e.g.][]{Wen09,Wen12,Szabo11} They are then confirmed by follow-up spectroscopy. 
They can also be identified by X-ray observations as extended sources, unambiguously testifying the presence of hot gas trapped in the potential well of a virialised  system \citep[e.g.][hereafter XXL paper I]{Pierre04,Pacaud07,Pierre16}. In comparison to optically selected samples, X-ray selected cluster samples 
are much less affected by projection effects, and their properties can be measured with good accuracy. In addition, the best way to detect active galaxies is through X-ray observations \citep[e.g.][]{Brandt15}.  

However, the effect of the group and cluster environment on the activity of the central supermassive black hole (SMBH) of galaxies
and vice versa is still fairly undetermined. Although it has been clearly established that an excess of X-ray point-like sources is found within clusters
\citep[e.g.][]{Cappi01,Molnar02,Johnson03,DElia04,Cappelluti05,Gilmour09}, recent studies have reported a significant lack of AGNs in rich galaxy clusters with respect to the field \citep[e.g.][]{Haines12,Ehlert13,Ehlert14,Koulouridis10}. Nevertheless, spectroscopic studies of X-ray point-like sources in rich galaxy clusters have concluded that low X-ray luminosity AGNs ($L_{\rm X}<3\times 10^{42}$erg s$^{-1}$) 
are equally present in cluster and field environments \citep[e.g.][]{Martini07,Haggard10}. In addition, most of the X-ray sources found in clusters present weaker optical AGN spectrum than AGNs in the field \citep{Marziani17} or show no signs of an optical AGN \citep[e.g.][]{Martini02,Martini06,Davis03}. However, luminous AGNs are rarely found in clusters 
\citep[][]{Kauffmann04,Popesso06a,Caglar17}.
 
On the other hand, although AGNs are not found in cluster cores \citep[e.g.][XXL paper XXI]{Sochting02,Ehlert14,Melnyk17}, there is evidence that X-ray AGNs found in massive clusters are an in-falling population located mostly in the outskirts \citep{Haines12}. Furthermore, \citet{Ehlert15} argues that an important part of the cluster AGN population is triggered by galaxy mergers. Theoretically, the feeding of the black hole can be enhanced by means of non-axisymmetric perturbations that induce mass inflow during galaxy interactions and merging.
This can lead to the AGN triggering \citep[e.g.][]{Kawakatu06,Koulouridis06,Koulouridis06b,Koulouridis13,Koulouridis14a,Ellison11,Silverman11,Villforth12,Hopkins11,Hopkins14}, rendering the high-density cluster surroundings a favourable
AGN environment. While this maybe the case for the outskirts, the rather extreme 
conditions within the innermost cluster regions seem to have the opposite effect. 
In more detail, the ram pressure stripping from the intracluster medium
(ICM) is probably able to strip or evaporate the cold gas reservoir of galaxies
\citep[][]{Gunn72,Cowie77,Giovanelli85,Popesso06a,Chung09,Jaffe15,Poggianti17b} and can strongly affect 
the fueling of the AGN. 
However, we note that the efficiency of ram pressure in transforming blue-sequence galaxies to red has been challenged
\citep[e.g.][]{Larson80,Balogh00,Balogh02,Bekki02,vandenBosch08,Wetzel12} and halo gas stripping has been suggested instead \citep[`strangulation'; e.g.][]{Larson80,Bekki02,Tanaka04}. 
Nevertheless, recent spectroscopic observations demonstrated 
the efficiency of ram pressure stripping in cluster galaxies \citep[][]{Poggianti17b}. Interestingly, in merging
or actively growing clusters the high incidence of galaxy mergers
can potentially enhance the number of AGNs, while at the same
time, shock waves in the ICM generated by cluster mergers may also enhance the ram pressure stripping \citep[e.g.][]{Vijayaraghavan13,Jaffe16}. 

Whichever the exact physical mechanism, if the suppression of accretion onto the SMBH is due to
the gravitational potential, we would expect the AGN presence to anti-correlate with cluster mass. However, 
the majority of previous studies have investigated rich X-ray clusters \citep[e.g.][]{Martini06,Koulouridis10,Haines12} or optically selected galaxy groups \citep[e.g.][]{Bitsakis15}. \citet{Arnold09} studied a sample of 16 local groups and rich clusters and found evidence of this anti-correlation. \citet{Oh14} reported similar findings for a sample of 16 poor X-ray clusters at intermediate redshifts. In \citet{Koulouridis14}, we investigated the AGN frequency in 33 poor and moderately rich clusters of the XMM-LSS survey up to $z=1$ and found no AGN suppression near the cluster centres, as would be expected for more massive structures \citep[e.g.][]{Ehlert13,Ehlert14,Koulouridis10}. In addition, \citet{Ehlert15} argued that the number of AGNs in a sample of 135 clusters scales with $M^{-1.2}$. On the other hand, studying the fraction of luminous AGNs ($L_{\rm X}>10^{43}$ erg sec$^{-1}$) in the DES cluster sample, \citet{Bufanda17} reported no differences between groups and clusters at any redshift. We note, 
however, that the differences in the sample mass range of the above studies may account for the discrepant results. For example, there is no cluster in the sample of \citet{Ehlert15} below the mass threshold of $10^{14} M_{\sun}$ which has been used in the literature to separate groups from clusters \citep{Bufanda17}.  

In the current work we 
study the X-ray AGN frequency in 167 spectroscopically confirmed X-ray selected clusters of the XXL survey (The Ultimate XMM-Newton Survey, see XXL paper I), spanning a redshift range of $z$=0.1--0.5. This homogeneous collection of relatively low-mass X-ray clusters provides a far larger sample of such clusters than previous studies. This mass range is largely unexplored, but is nevertheless an essential link between optical galaxy groups and massive clusters.  
With its depth, uniform coverage, and well-defined selection function,
the XXL Survey, is making a unique contribution to the study of clusters. More importantly, its two 25 deg$^2$ fields are essential to the study of AGNs in the cluster environment, since clusters can be 
very extended, of the order of a few Mpc, and AGNs may preferentially reside in their outskirts.  

In Sect. 2 of the paper we present the cluster and AGN catalogues. In sec. 3 we describe the applied methodology and in sec. 4 we present the results.  Finally, we draw our conclusions and discuss the results in sec. 5. Throughout this paper we use $H_0=70$ km s$^{-1}$ Mpc$^{-1}$, $\Omega_m=0.3$, and $\Omega_{\Lambda}=0.7$.

\section{Data description}
\subsection{The XXL Survey}

The XXL Survey is the largest {\it XMM-Newton} project approved to date ($>$6 Msec), 
surveying two $\sim$ 25 deg$^2$ fields, called the northern (XXL-N) and the southern (XXL-S) fields, with a median exposure of 10.4 ks 
and a depth of $\sim5\times10^{-15}$ erg sec$^{-1}$ cm$^{-2}$ in the (0.5--2 keV) soft X-ray band
(completeness limit for point-like sources). The two fields have extensive multiwavelength coverage 
from X-ray to radio. A general description of the survey and its goals was published in XXL paper I.
To date approximately 450 new galaxy clusters have been detected out to redshift $z\sim2$, and more than 20000 AGNs 
out to $z\sim4$. The main goal of the project is to constrain the dark energy equation 
of state parameter, $w$, using clusters of galaxies. This survey will also be a lasting legacy for cluster scaling laws 
and studies of galaxy clusters, AGNs, and X-ray background \citep[see also][]{Pierre17}.

\subsection{Galaxy cluster sample}
\begin{figure}
\centering
\resizebox{8cm}{12cm}{\includegraphics[angle=270, origin=c]{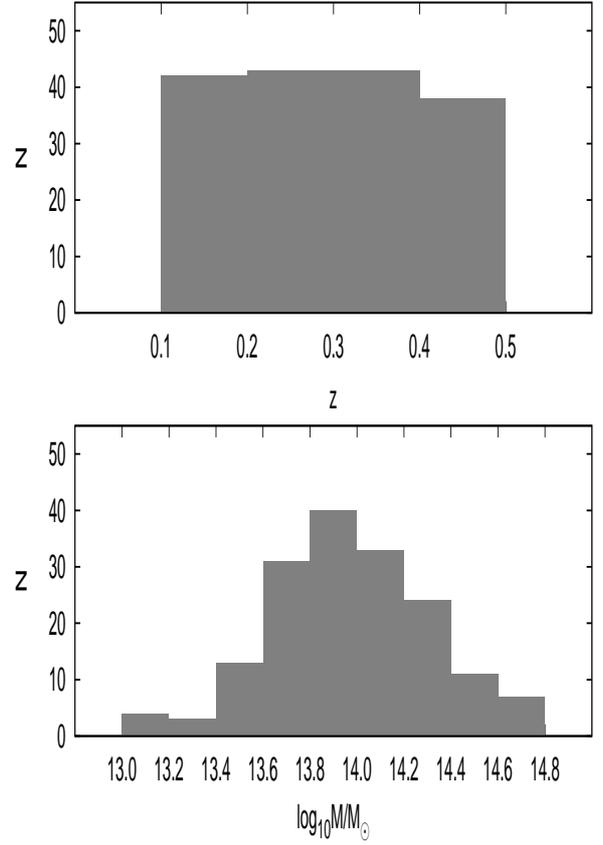}}
\caption{Redshift (top panel) and mass distribution (bottom panel) of the 167 spectroscopically confirmed C1 and C2 clusters in the two XXL fields.}
\label{fig:stats}
\end{figure}

From the total cluster catalogue of the XXL survey we select all spectroscopically confirmed class-1 (C1) and class-2 (C2) clusters spanning a redshift range of $0.1<z<0.5$. The lower redshift limit discards a small number of nearby clusters because their angular diameter is very extended, even for the 25 deg$^2$ of each XXL field. The upper redshift limit allows us to uniformly detect AGN cluster members down to a lower soft-band (0.5--2 keV) X-ray luminosity of $L_{\rm X [0.5-2\,keV]}>10^{42}$ erg cm$^{-1}$. In addition, we limit our sample to these clusters with direct X-ray temperature and luminosity measurements (see next paragraph for more details). The above selection yields 121 C1 and 46 C2 X-ray selected galaxy clusters. 

The C1 and C2 selection criteria are described in \citet{Pacaud06} and the properties of the clusters are thoroughly studied and presented in \citep[][hereafter XXL paper XX]{Adami18}. The C1 selection guarantees a pure cluster sample, while the C2 selection pertains to les massive clusters and is a priori contaminated by false detections (30-50\%  are point-like sources mistaken for clusters). Therefore, we selected only the spectroscopically confirmed clusters to avoid including false sources. The cluster mass and the $r_{500}$\footnote{Overdensity radius with respect to the critical density.} radius are well defined for our clusters, which is crucial for the current study. In more detail, we were able to measure directly the X-ray temperature and luminosity performing a spectral analysis of the X-ray observations. 
Initially, the extent  of  the emission was defined as the radius beyond which no significant cluster emission is detected using a threshold of 0.5$\sigma$ above the background level. For detailed background modelling we followed the methodology described in \citet{Eckert14}. The soft proton background was modelled with a broken power law, and the non X-ray background using closed filter observation. In addition, the  sky  background  was  modelled using data extracted from an offset region (outside the cluster emission), using a three-component model. Finally, cluster source spectra were extracted for each of the XMM-Newton cameras separately and the (0.4--11.0 keV) band was modelled with an absorbed Astrophysical Plasma Emission
Code  \citep[APEC;][]{Smith01} model (v2.0.2) with a fixed metal abundance of $Z=0.3Z_{\sun}$. Using the mass-temperature scaling relation of \citet[][XXL paper IV]{Lieu16} we 
calculated the ($M_{500,MT}$) mass and the $r_{500,MT}$ radius. The mass and redshift distribution of our sample is presented in Fig. \ref{fig:stats}.

Although there is no clear definition, galaxy concentrations more massive than $10^{14}M_{\sun}$ are defined as galaxy clusters, while less massive aggregations are called galaxy groups. According to the above classification almost half of the extended X-ray sources in the current study fall in the former category and the rest in the latter. Our sample includes very few clusters above a mass of $M_{500,MT}\sim3\times10^{14}M_{\sun}$. Our sources cover an estimated mass range of $10^{13}-5\times10^{14} M_{\sun}$, which classifies them as poor clusters (groups) or moderately rich clusters. This is an important feature of our sample for the current study, allowing us to investigate the role of the cluster mass in triggering and suppressing AGN activity.

\subsection{X-ray point-source sample and spectroscopic redshifts}

The X-ray point source catalogue is thoroughly described in \citet[][XXL paper XXVII]{Chiappetti18}, 
where all X-ray and the associated ancillary data (infrared, near-infrared, optical, ultraviolet, radio, and spectroscopic redshift when available) of 26056 X-ray sources are described. 

Spectroscopic redshifts were obtained with large spectroscopic surveys with which we have collaborative agreements, for example SDSS, VIPERS \citep{Guzzo14}, and GAMA \citep{Liske15} in  XXL-N, and from a large campaign with the AAOmega spectrograph on
the  Anglo-Australian  Telescope \citep[see][XXL paper XIV]{Lidman16}. Other smaller scale spectroscopic observations \citep[e.g. with WHT,][XXL paper XII]{Koulouridis16} complement the sample.

All spectroscopic information is hosted in the 
Centre  de  donnéeS  Astrophysiques  de  Marseille (CeSAM)  database  in
Marseille\footnote{http://www.lam.fr/cesam/}. The second data release
(CeSAM-DR2) is public and can be downloaded directly from
the database. 

\section{Methodology}

\subsection{Spatial density of optical galaxies}

Any excess of X-ray AGNs in the area of galaxy clusters can be due to the obvious abundance of galaxies with respect to the field
 \citep[e.g.][]{Koulouridis10,Haines12,Koulouridis14}. Therefore, to reach a meaningful interpretation of the X-ray AGN density we need to compare it to the density of optical sources. 
 To this end we have used the clusters of XXL-N that fall in the CFHTLS-W1 field and therefore have reliable photometric redshifts \citep[][]{Ilbert06,Coupon09}. The methodology used to select galaxies and define the background is described in detail in \citet[][XXL paper XXVIII]{Ricci18}. To summarise, we selected galaxies with photometric redshifts in a certain window around the spectroscopic redshift of each cluster. This window was precisely defined in order to obtain a homogeneous 68\% membership completeness that depends both on cluster redshift and galaxy magnitude. 
 In order to select bright galaxies in a homogeneous way, we used an evolutionary model as reference for the redshift evolution of the characteristic apparent
magnitude $m^*$. This model was computed with LePhare using
the elliptical galaxy SED template `burst\_sc86\_zo.sed' from the
PEGASE2 library \citep{Fioc97}, with a redshift  of formation $z_f=3$. We normalised the model using K* values from \citet{Lin06}, and corrected to AB system. This leads to a magnitude of $M^*_i=-21.75$ at $z=0$ in the $i'$ band, which does not significantly evolve within the redshift span of the current study. Finally, we selected galaxies (and counterparts of the X-ray sources) having an absolute magnitude $M_i$ in the i band within $M_i^*-2<M_i<M_i^*+1$.  
 The background galaxy density was computed as the mean galaxy density in the entire CFHTLS-W1 field.

 For the purposes of the current study we calculated two optical profiles, one for poor clusters with masses below $M_{500,MT}=10^{14} M_{\sun}$ and one above this threshold, since the mass-range of our clusters is rather narrow. The analysed sample comprises $\sim$50\% of the total cluster sample and we assumed that it is representative of the full XXL population. For the rest of the clusters, one of the two available optical profiles was adopted on the basis of their estimated mass.
 
\subsection{X-ray AGN selection}

\begin{figure}
\resizebox{8cm}{7cm}{\includegraphics[angle=0]{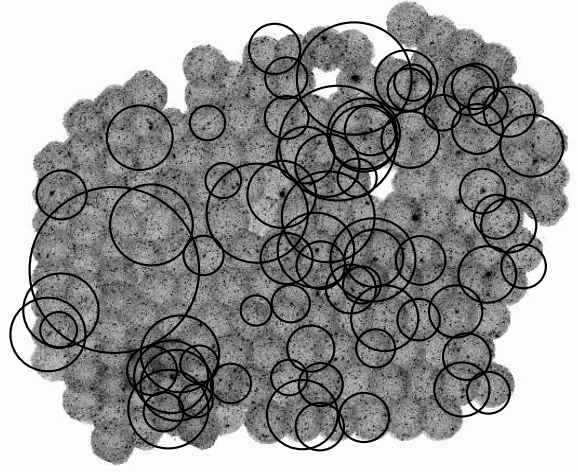}}
\caption{Position of the 75 spectroscopically confirmed X-ray selected clusters in the XXL-S field. The size of the circles in this figure correspond to 6$r_{500,MT}$ radius.}
\label{fig:map}
\end{figure}

\begin{figure*}
\centering
\resizebox{16cm}{12cm}{\includegraphics[angle=0]{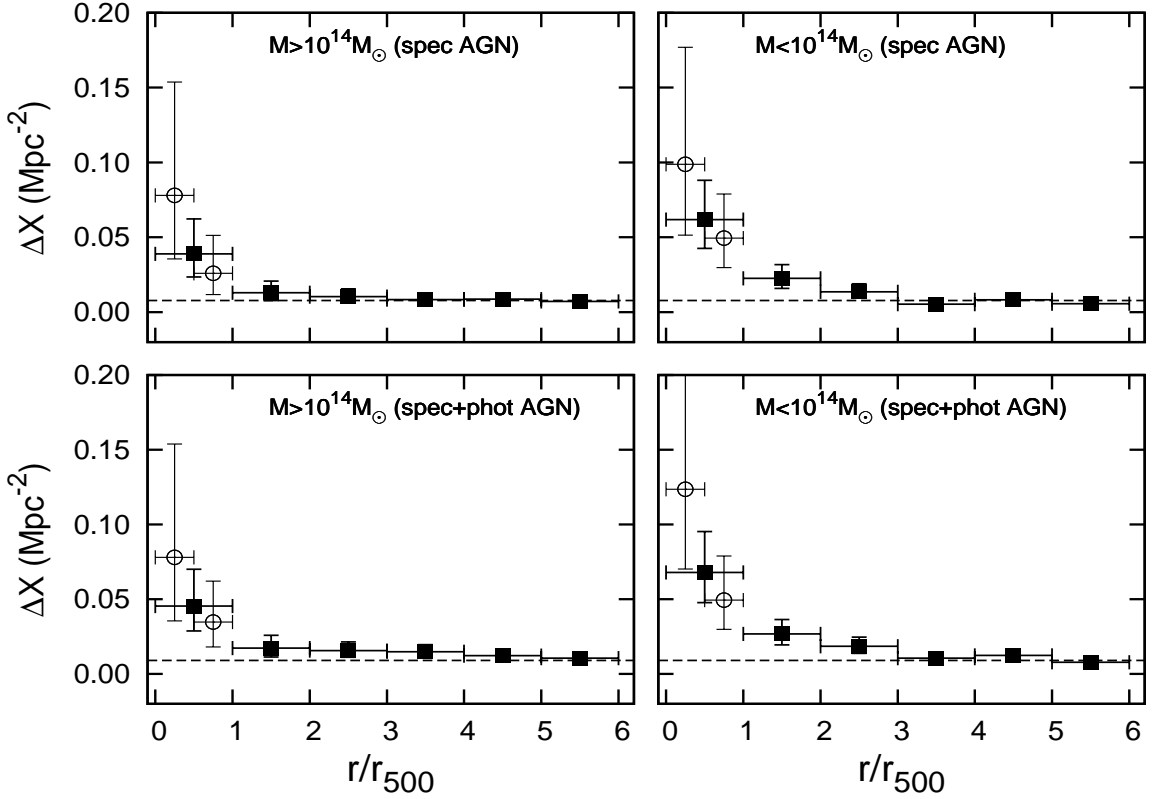}}
\caption{Density profiles of AGN cluster members, separated in two cluster mass bins (left and right panels). Only AGNs with spectroscopic redshifts are included in the top panels, while photometric redshifts are also used in the bottom panels. The dashed line represents the AGN field density as calculated from 100 mock catalogues in each case. Open circles mark the results of the first  bin split into two. Error bars indicate the 1$\sigma$ confidence limits for small numbers of events \citep{Gehrels86}.}
\label{fig:prof}
\end{figure*}

In our analysis we considered all X-ray point-like sources that (a) fall within $6r_{500,MT}$ of the centre of each cluster (see Fig. \ref{fig:map}), (b) have X-ray luminosities in excess of $L_{\rm X [0.5-10\,keV]}>10^{42}$ erg sec$^{-1}$, and (c) whose optical counterparts have absolute i-band magnitudes within $M_i^*-2<M_i<M_i^*+1$, consistently with the range used for optical galaxies in Sec. 3.1. Of the above sources, $\sim$90\% and $\sim$70\% have a spectroscopic redshift in XXL-S and XXL-N, respectively. We divided the area around each cluster into six annuli of $r{500}$ radius and we counted all the sources that fell into one of the following categories: spectroscopic members, for which the maximum radial velocity difference between the galaxy and the cluster is chosen at $\Delta u=\pm$2000($z_{spec}+1)$ km/sec \citep[e.g.][]{Koulouridis10, Koulouridis14, Koulouridis16, Martini13} or photometric members, which are galaxies with low probability of being stars or outliers, and with a narrow probability distribution function (PDF) around the cluster 
redshift. The first annulus is further divided because the detection of X-ray sources close to the cluster core may be affected by the extended X-ray emission of the hot gas. Therefore, the number counts in the innermost region is always a lower limit. We also note that X-ray AGNs associated with the brighter cluster galaxies were removed, since their triggering and evolution is not the subject of the present study. 

To compute the background level we constructed 200 mock catalogues, 100 for each XXL field. We match the number and redshift distribution of our clusters, randomising only their position in the fields. The most important property of the mock catalogues is that they follow the X-ray sensitivity of the survey allowing us to tackle selection effects. The average density of X-ray sources was found to be roughly the same in all annuli. 

In Fig. \ref{fig:prof} we present the X-ray AGN density profiles in the two fields using both selections.  The profiles were constructed by computing the density of X-ray AGNs, $\Delta X$, in each annulus $i$, following the formula \[\Delta X_{i}=\sum_1^{N_c} N_i/\sum_1^{N_c} A_i,\] 
where $N_c$ is the number of clusters, $N_i$ is the number of X-ray AGNs in annulus $i$, and $A_i$ the respective area.
The differences between using all AGNs or only the spectroscopically confirmed ones are not significant. Considering that the spectroscopic completeness of our sample is high (especially in  XXL-S), while on the other hand AGN photometric redshifts have large uncertainties, we chose to proceed using only the spectroscopically confirmed members.

For the presentation of the results we computed the AGN fraction in cluster member galaxies, or the ratio of cluster AGN fraction to the field AGN fraction. Our results are compared with similar studies of the AGN fraction in bright cluster  galaxies selected in various bands. The majority of these studies use an $M^*+1$ magnitude cut below the knee of the luminosity function, consistent with that used in the current study. For the analyses that use $M^*+1.5$, \cite{Martini13} argued that any possible discrepancy caused by the 0.5 magnitude difference is far smaller than the Poisson uncertainties. Errors were computed based on the confidence limits for small number of events \citep{Gehrels86}.          
 
\section{Results}

\begin{figure}
\resizebox{9cm}{7cm}{\includegraphics[angle=0]{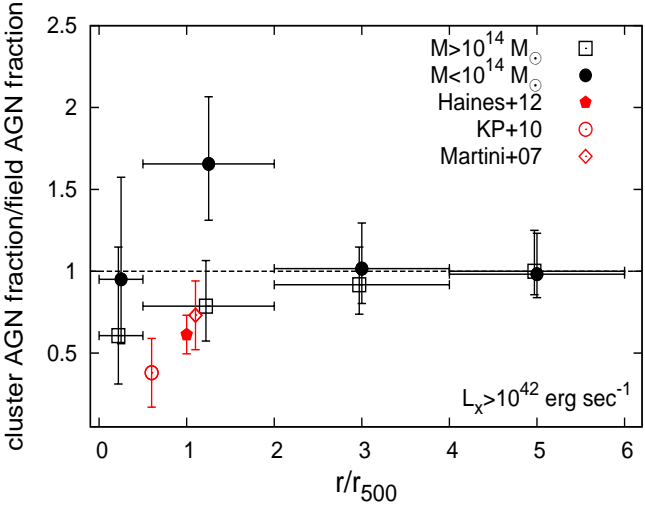}}
\caption{Fraction of XXL bright cluster galaxies hosting an X-ray AGN ($L_{\rm X [0.5-10\,keV]}>10^{42}$ erg sec$^{-1}$), divided by the field fraction. The results are plotted as a function of distance from the cluster centre. The sample is divided in two based on cluster mass. Error bars indicate the 1$\sigma$ confidence limits for small numbers of events \citep{Gehrels86}. For comparison we plot results from the analyses of massive clusters by \citet[][8 clusters, 0.06$<z<$0.31]{Martini07}, \citet[][16 clusters, 0.07$<z<$0.28]{Koulouridis10}, and \citet[][26 clusters, 0.15$<z<$0.30]{Haines12}. A significant AGN excess is found between 0.5$r_{500,MT}$ and 2$r_{500,MT}$, at the 95\% confidence level.}
\label{fig:over}
\end{figure}

In Fig. \ref{fig:over} we present the results of our analysis, where we plot the fraction of X-ray AGNs in cluster galaxies to the fraction of X-ray AGNs in field galaxies up to $6r_{500}$ radii. We divided the cluster population into low- and high-mass structures using the $M_{500,MT}=10^{14} M_{\sun}$ threshold. This limit is frequently used in the literature to separate groups from massive clusters \citep[e.g.][]{Bufanda17}. The two subsamples  roughly contain equal numbers of cluster galaxies.
To obtain statistically significant results we merge the two XXL fields.

The first important result is the difference between the low- and high-mass cluster samples. A significant excess of X-ray AGNs, at the 95\% confidence level, is found in the first $2r_{500,MT}$ range of the low-mass cluster population. In contrast, the AGN fraction in the high-mass sample is consistent with the field value, and there is also evidence of a decreasing trend towards the cluster centres. The results of the innermost $0.5r_{500,MT}$ annulus are plotted separately since they may be affected by observational effects (see Sect. 3.2). The number of AGNs detected in cluster cores is small and the statistics are poor, but we find a decreasing trend, especially for the low-mass sample in which the AGN fraction falls to the field level.

We also compare our results with those of massive cluster studies. In Fig. \ref{fig:over} we overplot the results of 16 massive Abell clusters presented in \citet{Koulouridis10}. Spectroscopic redshifts were not available and therefore the projected X-ray AGN densities were computed, statistically subtracting the background. Data were available within a constant radius of 1 Mpc for most of the sources, which is close to the $r_{500}$ radius for rich clusters. Optical galaxy densities were computed in the same regions respectively, within $m^*_r-0.5<m_r<m^*_r+0.5$ (SDSS). We also plot the results of a spectroscopic study of 26 massive clusters that reported 40\% fewer X-ray AGNs in galaxies brighter than $M^*_K+1.5$  within 2$r_{500}$ \citep{Haines12} when compared to the field density \citep{Haggard10}. They also found that these AGNs comprise an in-falling population. \citet{Martini07} found a similarly low AGN fraction in $M_R<-20$ galaxies for a smaller sample of seven clusters. The cluster-centric radius probed by the {\it Chandra} field of view varied from $r_{200}$ down to $\sim$0.5 Mpc. In sharp contrast to \citet{Haines12}, they argue 
that their AGN sample is not in-falling. Interestingly, their X-ray sources are systematically less luminous than the sample in \citet{Haines12}, although both studies  probe roughly the same galaxy population, up to $\sim M^*+1.5$. In addition, \citet{Martini07} reported that the most luminous AGNs ($L_{\rm X}>10^{42}$ erg sec$^{-1}$) are more centrally concentrated. This excess is significant and peaks at $\sim$0.5 Mpc. These results are consistent with our findings for the high-mass cluster population, but again is in contrast to the results of the low-mass sample. 

\begin{figure}
\resizebox{9cm}{7cm}{\includegraphics[angle=0]{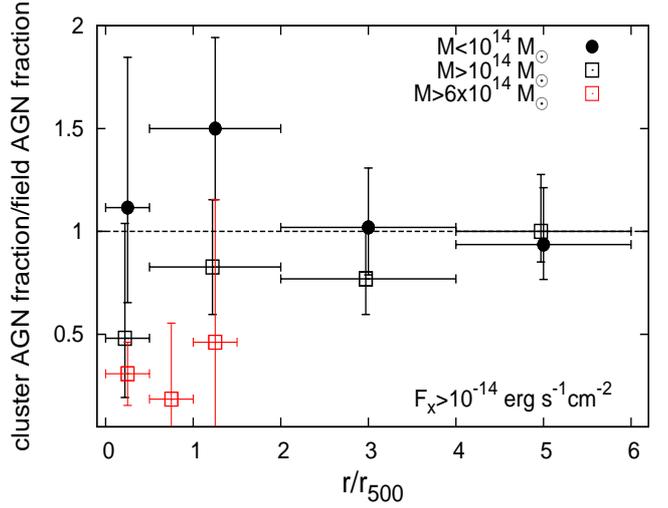}}
\caption{Fraction of XXL bright cluster galaxies hosting an X-ray AGN for a flux-limited sample ($F_{\rm X [0.5-10\,keV]}>10^{-14}$ erg sec$^{-1}$ cm$^{-2}$), divided by the field fraction. The results are plotted as a function of distance from the cluster centre. No AGN suppression is present in our low-mass cluster sample, in contrast to the results of \citet{Ehlert14} on massive clusters. Error bars indicate the 1$\sigma$ confidence limits for small numbers of events \citep{Gehrels86}.}
\label{fig:over2}
\end{figure}

Furthermore, \citet{Ehlert14} used a sample of 42 massive clusters ($z<0.7$) to study the X-ray AGN fraction in galaxies with SuprimeCam $R$-band apparent magnitudes brighter than $R$=23. Their least massive cluster ($\sim6\times 10^{14} M_{\sun}$) is more massive than our most massive cluster. However, they used a lower flux limit of $10^{-14}$ erg sec$^{-1}$ cm$^{-2}$, selecting only the brightest X-ray sources.  We also note that spectroscopic redshifts were not available. For a meaningful comparison we apply the same flux limit to our samples and we plot the results in Fig. \ref{fig:over2}. Similarly to Fig. \ref{fig:over}, an AGN deficiency is found in both high-mass samples, while the excess is only present in the low-mass sample.

To test the stability of our results,  
we  repeated the analysis used in  this paper using an alternative cluster mass estimation. In more detail, we performed aperture photometry within 300 kpc in the (0.5--2 keV) band from the cluster centre. Then we followed an iterative procedure using the M-L and the L-T scaling relations, fully described in XXL paper XX, and we obtain the respective $M_{500,scal}$ and $r_{500,scal}$ values. This method allows the use of the full XXL sample of 209 spectroscopically confirmed clusters between $z=0.1$ and 0.5. 
 Qualitatively, the results of the present study are not affected and the conclusions remain the same. We note, however, that this method is less precise than the one using direct temperature measurements that we have used in the current study.

\begin{figure}
\resizebox{9cm}{7cm}{\includegraphics[angle=0]{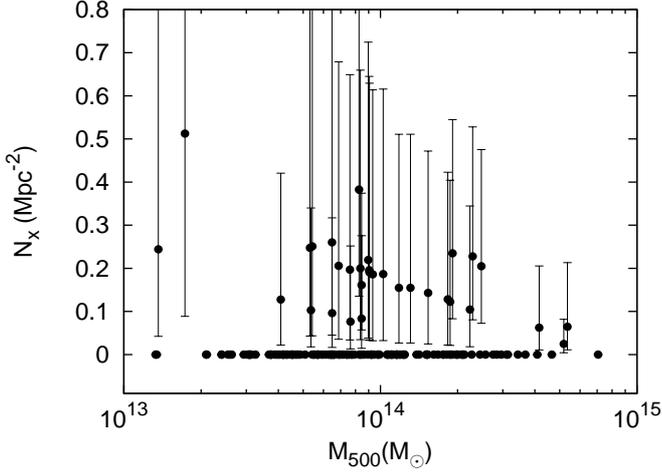}}
\caption{AGN density per cluster within $2r_{500}$ as a function of cluster mass. Error bars indicate the 1$\sigma$ confidence limits for small numbers of events \citep{Gehrels86}.}
\label{fig:indiv}
\end{figure}

Finally, in Fig. \ref{fig:indiv} we plot the density of AGN per cluster within $2r_{500,MT}$ as a function of cluster mass. These results are similar to those presented in Fig. 6 of \citet{Ehlert15}. In both cases, the statistical significance of AGN excess in individual clusters is small, while several clusters do not contain any AGNs. This is due to the rarity of X-ray AGNs in the total galaxy population. As a result, the expected number density of X-ray AGNs above the luminosity threshold per cluster is very low. Therefore, very rarely is more than one AGN  detected within a $2r_{500}$ radius of an individual cluster. However, there is evidence of increasing densities towards less massive clusters. On the other hand, this trend also reflects the declining $r_{500}$ radius in the same direction. Therefore, we argue that in order to obtain reliable results it is necessary to stack the AGN number counts, as described in the current paper and in the literature \citep[e.g.][]{Koulouridis14,Ehlert15,Bufanda17}.

\section{Conclusions and discussion}

We studied the X-ray AGN population of a sample of 167 spectroscopically confirmed X-ray clusters. Our clusters cover the largely unexplored mass range of poor and moderately rich clusters ($10^{13}$- $4\times10^{14} M_{\sun}$). In this respect, the XXL sample is unique and even the next X-ray mission, e-Rosita, will not cover this mass range \citep{Borm14}.

Our results show a significant excess of X-ray AGNs in low-mass clusters ($M_{500,MT}<10^{14} M_{\sun}$), at the 95\% confidence level, up to a radius of $2r_{500,MT}$. Nevertheless, in the innermost cluster region ($<0.5r_{500,MT}$) a sharp decrease drives the AGN fraction back to the field level. On the other hand, the high-mass subsample presents a gradually decreasing AGN fraction towards the cluster centres, which reaches  a factor of three lower than in the field level. This result is in agreement with previous studies of AGNs in massive cluster samples. 

\begin{figure*}
\centering
\includegraphics[width=3in]{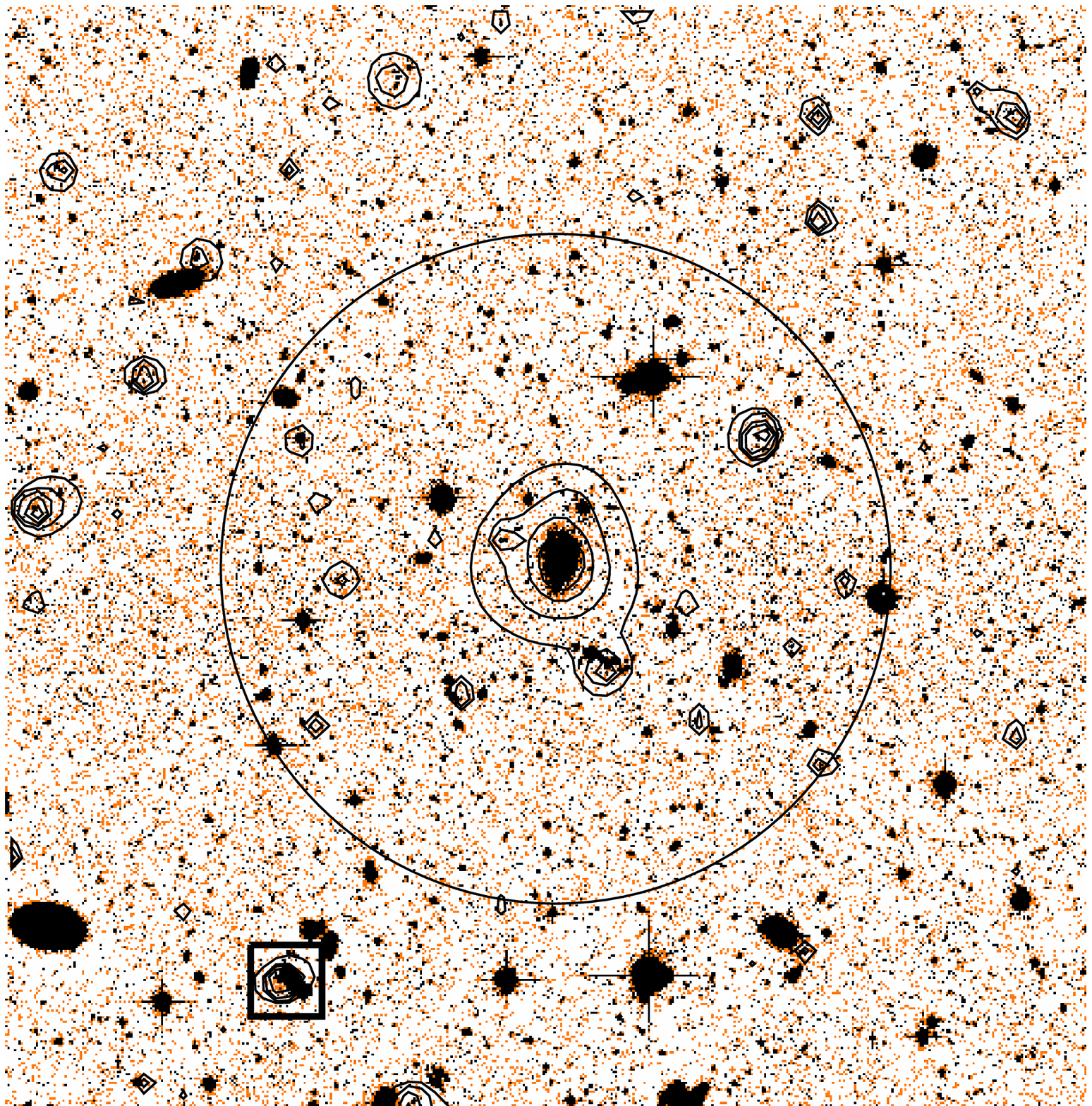}\hspace{2cm}\includegraphics[width=3.18in]{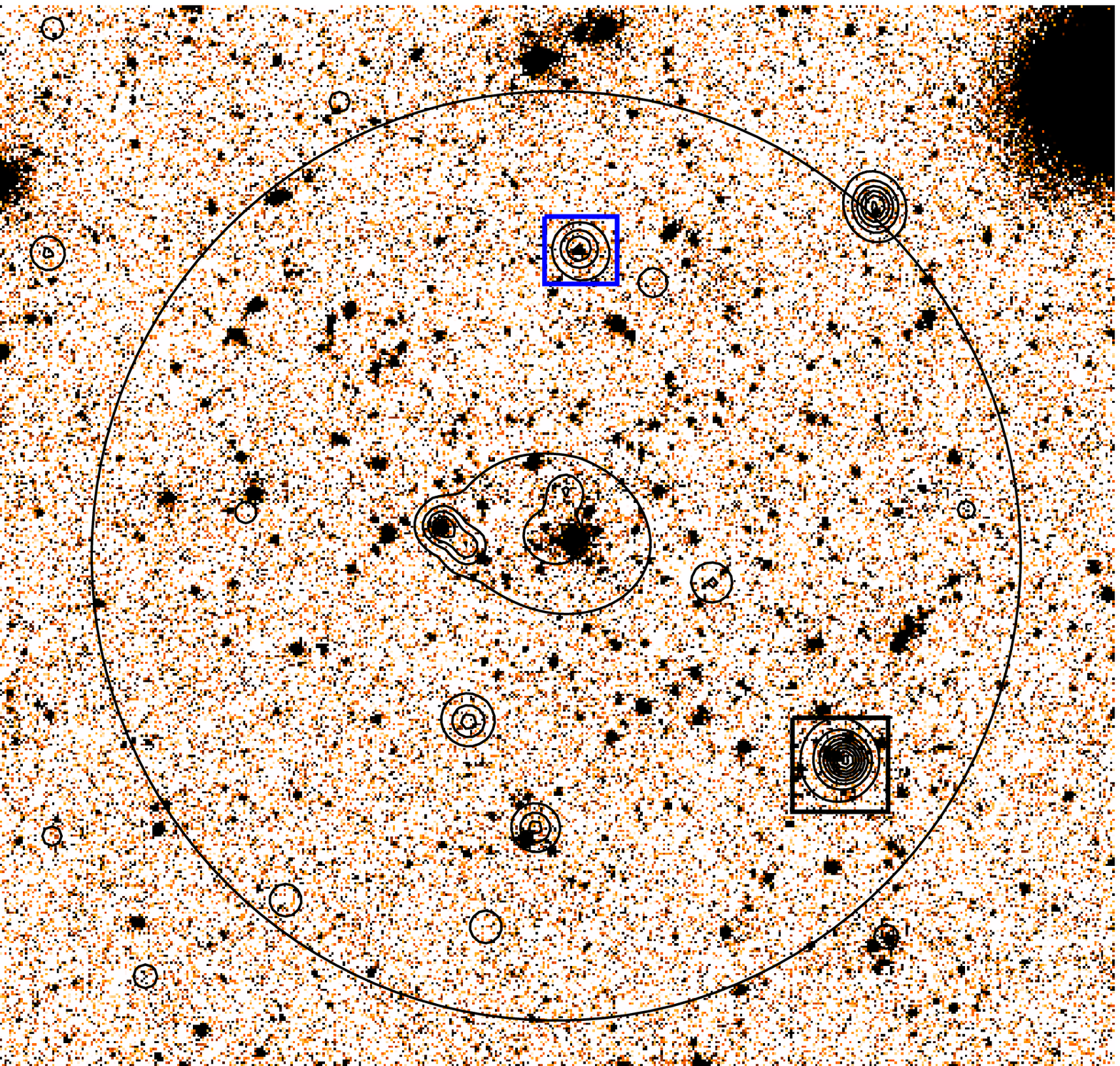}
\vspace{0.5cm}

\resizebox{16cm}{4cm}{\includegraphics[angle=0]{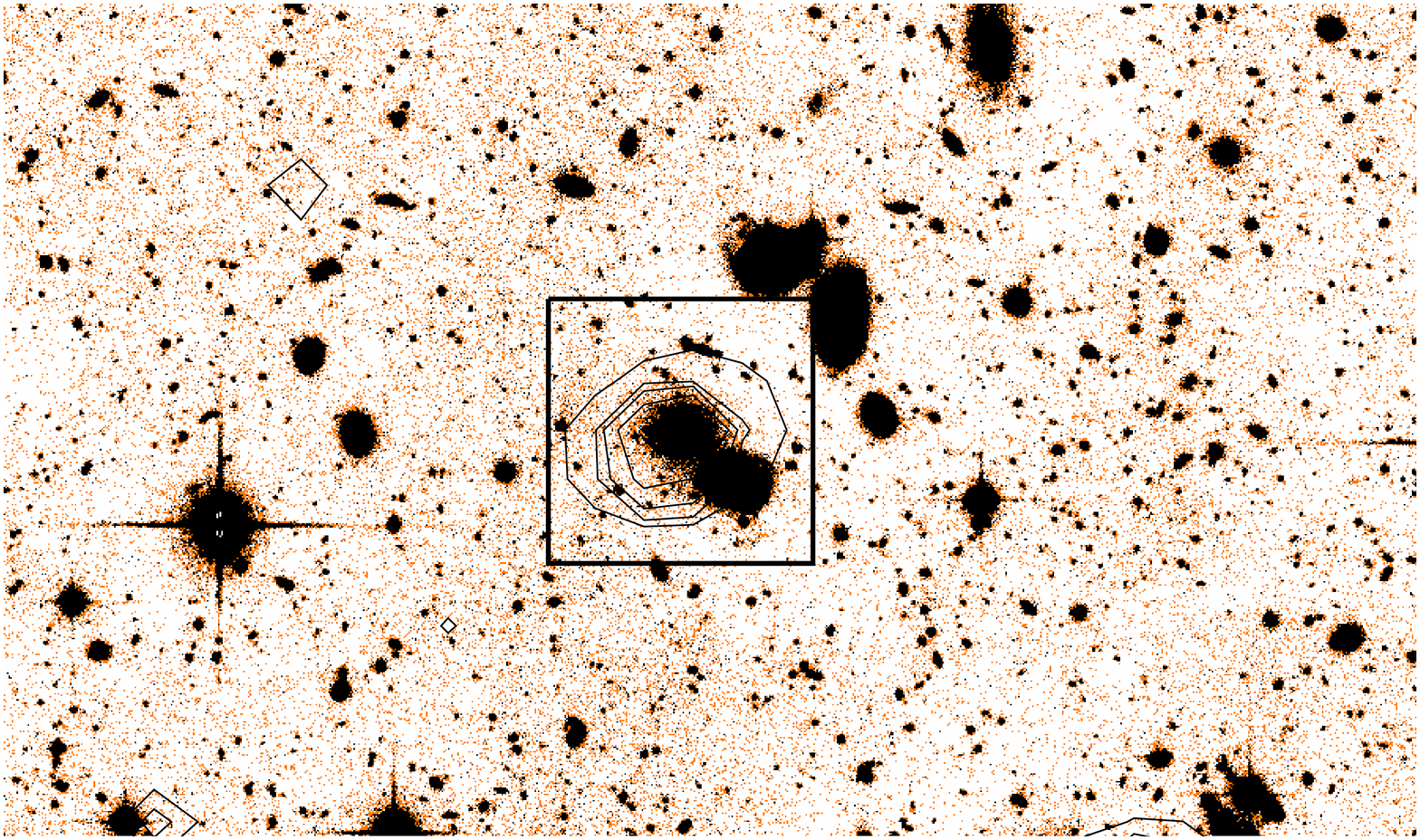}\hspace{22cm}\includegraphics{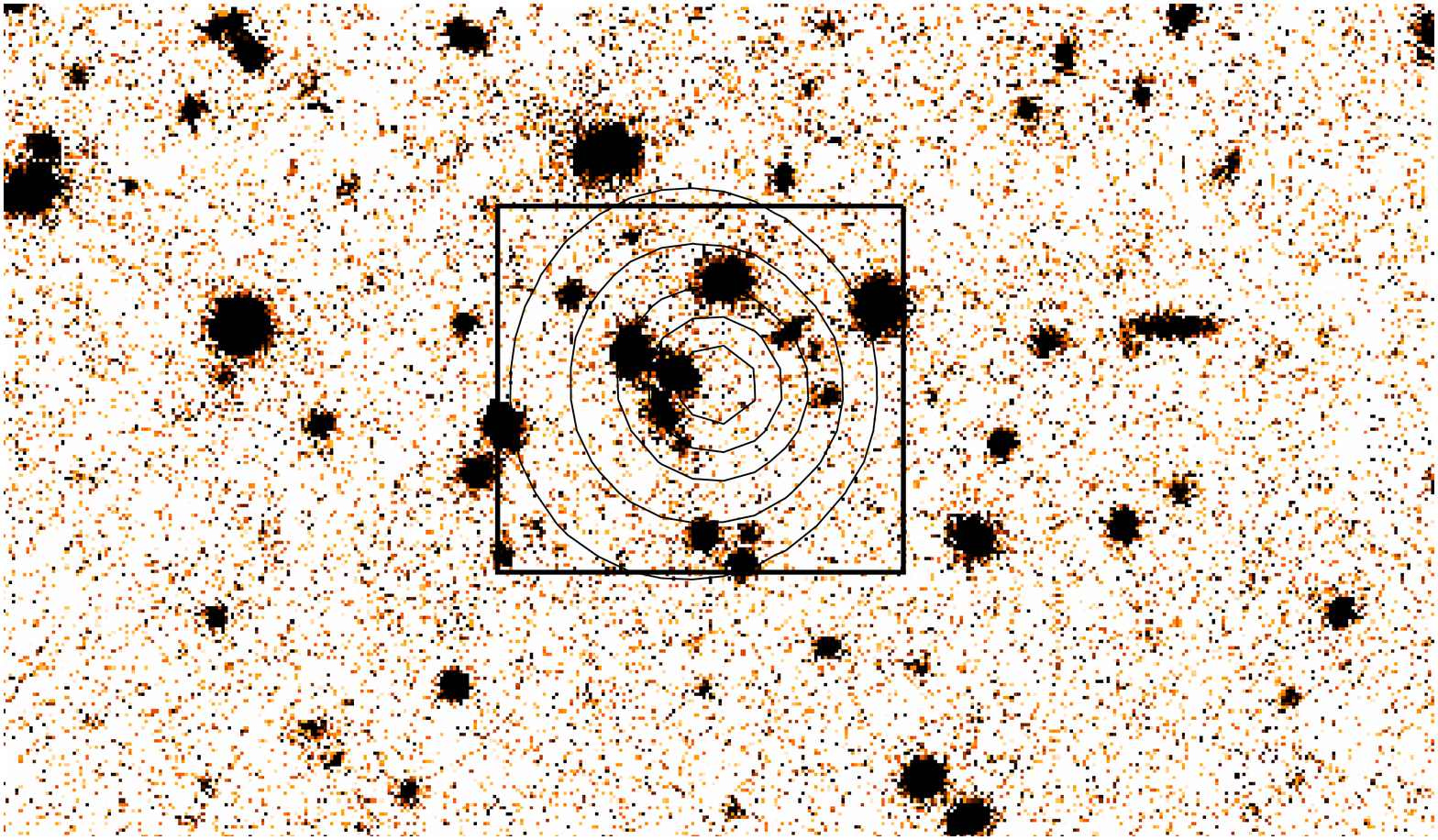}}
\caption{Optical images (i band) of two XXL clusters overplotted with X-ray contours (0.5--2.0 keV). Left panels: XLSSC 041 (in XXL-N), a C1 cluster at $z=0.142$ with an estimated mass of $M_{500,MT}\sim 8.3\times10^{13} M_{\sun}$ and $r_{500,MT}=636$ kpc. Right panels: XLSSC 561 (in XXL-S) a C2 cluster at $z=0.455$ with an estimated mass of $M_{500,MT}\sim 9.3\times10^{13} M_{\sun}$ and $r_{500,MT}=993$ kpc. The boxes indicate the position of AGNs with spectroscopic redshifts concordant with the respective cluster redshift. The bottom panels are zoomed-in images of the AGNs found in the black boxes. Both these AGNs are found in overdense regions, where galaxy interactions and merging is probable.}
\label{fig:optical}
\end{figure*}

We argue that the AGN excess found in the cluster outskirts of low-mass clusters in this study supports an in-falling population scenario. There is indeed evidence that AGNs can be found in surplus in the periphery of clusters \citep[e.g.][]{Johnson03,Koulouridis14}, and they were further shown to be an in-falling population in \citet{Haines12}. However, other studies claimed the opposite for low-luminosity AGNs \citep{Martini07}. The excess may be attributed to AGN triggering because of the higher galaxy density, which can lead to higher merger rates \citep{Ehlert15}. In Fig. \ref{fig:optical} we present two examples of X-ray clusters with spectroscopically confirmed AGN members in the first two annuli; galaxy interactions or merging is possible in both cases. However, high-velocity dispersions in massive clusters may diminish the probability of galaxy interactions. \citet{Arnold09} found that ten groups with velocity dispersion $\sigma<500$ km sec$^{-1}$ present higher AGN fractions than six clusters above 
this velocity dispersion limit. They used both X-ray ($L_{\rm X}>10^{41}$ erg sec$^{-1}$) and optically detected AGNs. 

We also argue that the decrease in the AGN fraction in the cluster cores, both in low- and high-mass clusters, supports the ram pressure stripping scheme produced by the hot intracluster medium. The hotter gas in deep gravitational potentials is expected to strip the galaxy more effectively than the colder gas in our poor cluster population. Indeed, the density of AGNs within the first $<0.5r_{500,MT}$ is consistent with the background level, in contrast to the well-established lack of AGNs in massive clusters.

Both of the  mechanisms described above lead to the differences between the AGN activity in poor and massive galaxy clusters. Previous results vary depending on the AGN selection, the cluster-mass range, and the selected cluster-centric radius. However, \citet{Ehlert15} also reported evidence of this cluster mass--AGN activity anti-correlation in a large sample of 135 clusters ($2\times10^{-14} - 2\times10^{-15} M_{\sun}$). They showed that the density of cluster AGNs has a strong dependence on cluster mass by modelling the behaviour of the individual projected X-ray point source density profiles of the clusters. We argue that the AGN dependence on cluster mass can only be investigated with large cluster samples in a statistical way. The AGN fraction in individual clusters varies considerably \citep[e.g.][]{Martini07}, and in most cases there are no AGNs detected. 
    
Because of the stripping, AGN activity should be weak in general. Early studies reported that many X-ray selected AGNs in clusters do not even present optical AGN signatures \citep{Martini02,Martini06}. In addition, \citet{Marziani17} provided strong evidence that optical emission is also weaker when compared to field AGNs. There is some evidence in our sample as well, where most of the detected AGNs present no broad Balmer lines in their optical spectra. In addition, some of them do not show any emission lines, as reported previously. However, no significant difference was found between the median X-ray luminosities of AGNs in the outer and inner annuli of the present study. We reanalysed our data including only X-ray AGNs with $L_{\rm X [0.5-10\,keV]}>5\times10^{42}$ erg sec$^{-1}$. No AGN deficiency was found in the first annuli that would suggest a higher density of low-luminosity AGNs. We repeated with $L_{\
rm X [0.5-10\,keV]}
>10^{43}$ erg sec$^{-1}$ and obtained similar results, although the number of detected sources is drastically reduced. However, we cannot investigate whether the excess of low-luminosity sources reported in previous studies is due to sources with $L_{\rm X}<10^{42}$ erg sec$^{-1}$. We do not consider these sources in our study because the XXL survey is shallow, and also because below this luminosity threshold the X-ray emission may be produced by alternative physical mechanisms (e.g. star formation, X-ray binaries). 

We  note that the low luminosities of X-ray sources may also be attributed to nuclear obscuration \citep[e.g.][]{Johnson03,Martini07}. High obscuration values may as well be due to recent galaxy merging or interactions \citep[e.g.][]{Hopkins08,Koulouridis14a,Koulouridis16a}.   
The activity and the type of our sources will be thoroughly explored in a future study. Nevertheless, a first approach showed that within the first two $r_{500}$ annuli the percentage of hard-band only detected AGNs, which imply obscuration, reaches 30\%, while it is $\sim$16\% in the three outer annuli.

\begin{figure}
\resizebox{9cm}{14cm}{\includegraphics[angle=0]{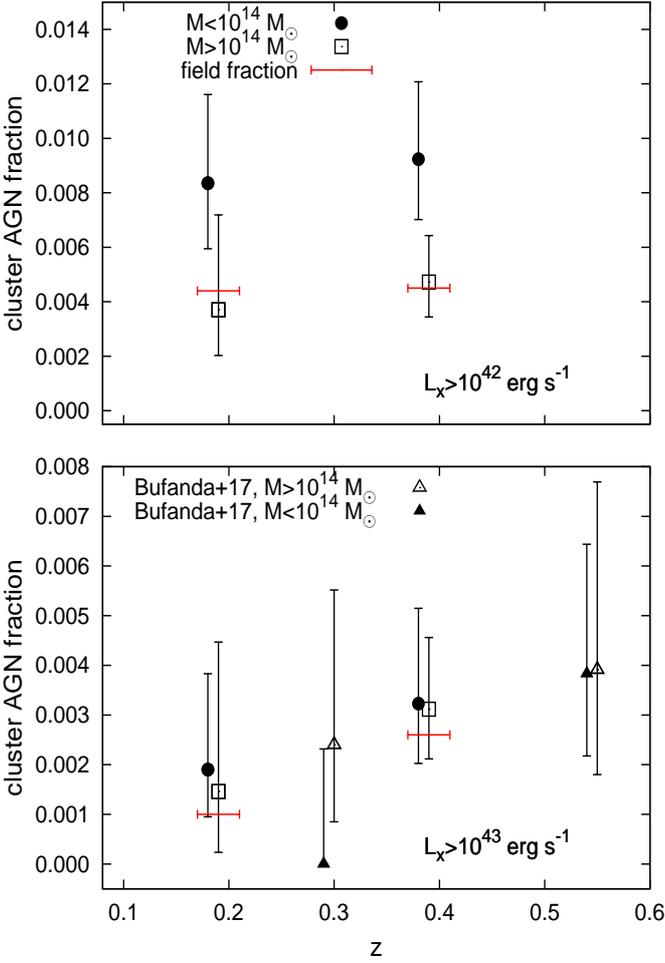}}
\caption{Fraction of XXL bright cluster galaxies hosting an X-ray AGN (top panel: $L_{\rm X [0.5-10\,keV]}>10^{42}$ erg sec$^{-1}$; bottom panel: $L_{\rm X [0.5-10\,keV]}>10^{43}$ erg sec$^{-1}$) within a $2r_{500}$ radius as a function of redshift. We overplot the results of similar analysis by \citet{Bufanda17}. Red lines indicate the respective field AGN fractions. Error bars indicate the 1$\sigma$ confidence limits for small numbers of events \citep{Gehrels86}.}
\label{fig:bufa}
\end{figure}

\begin{figure}
\resizebox{9cm}{7cm}{\includegraphics[angle=0]{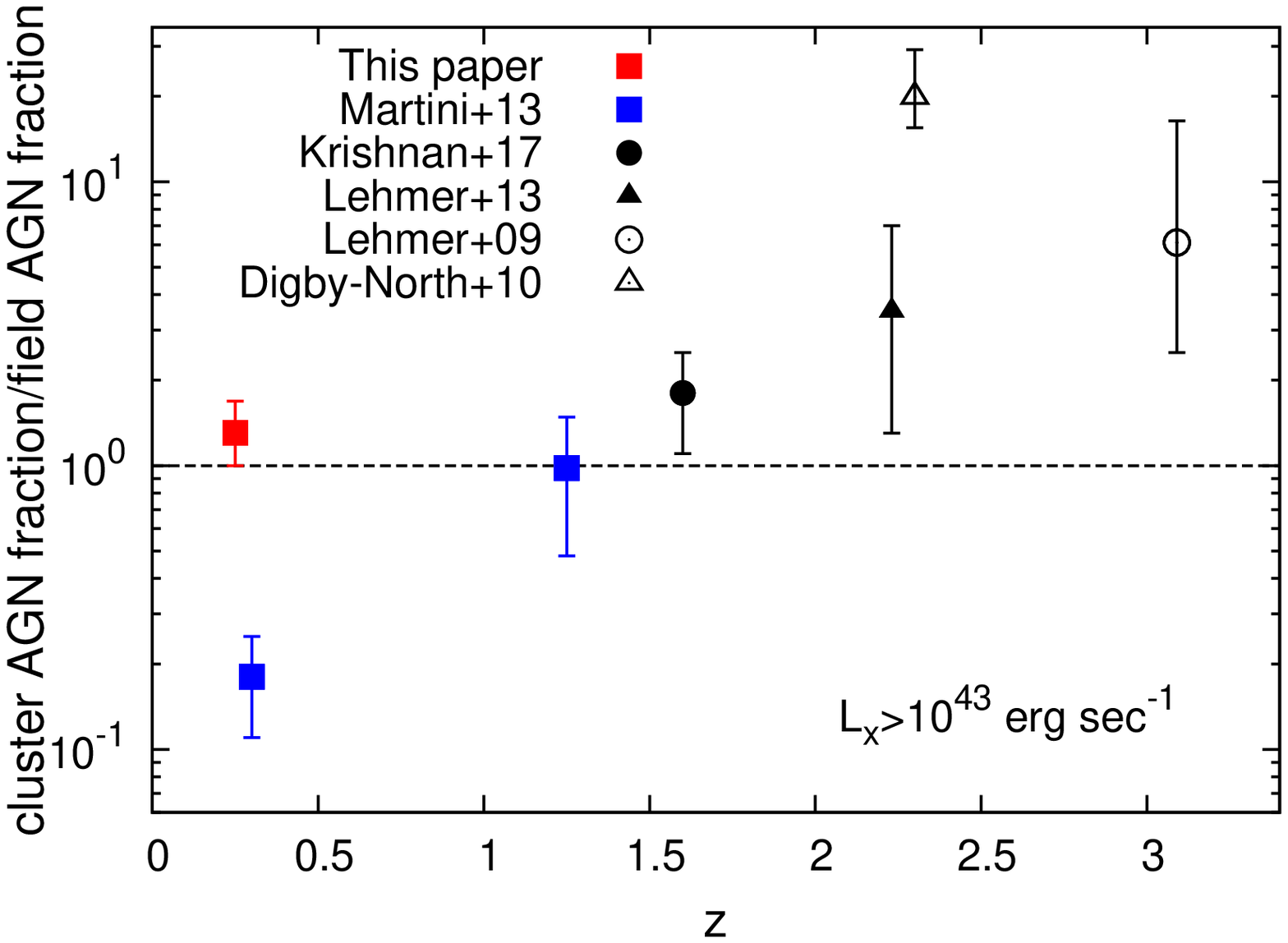}}
\caption{Fraction of XXL bright cluster galaxies hosting an X-ray AGN ($L_{\rm X [0.5-10\,keV]}>10^{43}$ erg sec$^{-1}$) within a $2r_{500}$ radius, divided by the field fraction. The results are plotted as a function of redshift. We overplot the results of similar analyses of cluster samples (blue) or individual high-redshift proto-clusters (black), as described in Sec. 5. Open symbols mark different luminosity limits used for the X-ray AGNs.}
\label{fig:highz}
\end{figure}

Interestingly, ram pressure stripping was also reported as a possible triggering mechanism for  AGN activity in cluster members \citep{Poggianti17a}, although the statistics are low (five sources), and all except one would not be selected by most X-ray AGN studies, being less luminous than $10^{42}$ erg sec$^{-1}$. Both effects of the intracluster gas seem plausible. {\it MUSE} spectroscopy of a jellyfish galaxy entering a cluster \citep{Fumagalli14} showed that despite the almost total stripping of the galaxy the core still retained a gas reservoir. Nevertheless, the core presented evidence of stripping as well. Therefore, initial pressure may indeed lead gas toward the galaxy core, triggering the nucleus, but eventually leads to strangulation.  

To investigate whether there is any evolution of the AGN fraction in cluster galaxies within our redshift range, we further divided our cluster population into two redshift bins, $z=$0.10--0.28 and $z=$0.28--0.50. The time interval in both bins is roughly equal to 1.9 Gyr. In Fig. \ref{fig:bufa} we plot the AGN fraction in clusters for the low-mass and high-mass subsamples separately. We also overplot the field value as calculated from our mock catalogues. When all sources above $L_{\rm X [0.5-10\,keV]}=10^{42}$ erg sec$^{-1}$ are considered (top panel) a slight evolution might be present, while the fraction of more luminous sources (bottom panel), $L_{\rm X [0.5-10\,keV]}>10^{43}$ erg sec$^{-1}$, clearly evolves more rapidly. The trend is consistent with the AGN fractions reported in \citet[][]{Bufanda17} in cluster galaxies with $M_R<M^*_R+1$, also plotted in Fig. \ref{fig:bufa}. Nevertheless, the AGN fraction in clusters evolves at the same rate as the AGN fraction in the field. Therefore, the ratio of the cluster to the field AGN fraction remains constant within the redshift range of the current study. Therefore, we conclude that any evolution of the cluster AGN fraction up to $z=0.5$ is not a result of the environment. Another interesting conclusion that can be drawn from this figure is that the AGN excess in low-mass clusters is produced by low-luminosity AGNs with $L_{\rm X [0.5-10\,keV]}<10^{43}$ erg sec$^{-1}$, equally present in both redshift bins. Similar results were also presented in \citet{Bufanda17}.

For a meaningful comparison with similar studies in high-redshift cluster samples and proto-clusters, we computed the fraction of AGNs with $L_{\rm X [0.5-10\,keV]}>10^{43}$ erg sec$^{-1}$ within  $2r_{500,MT}$ of the XXL cluster sample. In Fig. \ref{fig:highz} we plot our results and we compare them with two samples of low- and high-redshift clusters ($z$=0.3 and 1.3, respectively) from \citet[][$M_{R}<M^*_{R}+1$ and $M_{3.6}<M^*_{3.6}+1$, respectively]{Martini13}, a proto-cluster at $z$=1.6 \citep[][$M_*>10^{10} M_{\sun}$]{Krishnan17}, and a candidate proto-cluster at $z$=2.23 \citep[][H$\alpha$ emitting galaxies]{Lehmer13}. Two more proto-clusters at redshift $z=2.3$ \citep[][Lyman-break galaxies]{Digby-North10} and $z=3.09$ \citep[][mean AGN fraction among Lyman-break and Ly $\alpha$ galaxies]{Lehmer09} are plotted as well, although the AGN X-ray luminosity limits are higher. We note that proto-cluster studies above redshift 2 do not trace the same galaxy population since the methods for selecting cluster members are biased towards strongly star-forming galaxies \citep[for more details, see][]{Krishnan17}. Nevertheless, normalising all AGN fractions to their respective field fractions renders them more appropriate for an instructive comparison. In particular, the 13 clusters between $z$=1 and 1.5 in \citet{Martini13} have estimated masses $M>10^{14} M_{\sun}$ and up to a few times $10^{14} M_{\sun}$, consistent with our own high-mass sample. However, their low-redshift sample is much more massive than this, being the compilation of the sample of \citet{Haines12} and \citet{Martini09}. On the other hand, the mass of the proto-cluster at $z$=1.6 has an estimated mass of $5.7\times10^{13} M_{\sun}$, well bellow that of massive clusters in the local universe, but within the range of our low-mass X-ray clusters. We note that low-luminosity AGNs, which mainly produce the AGN excess in our cluster samples, are not included in Fig. \ref{fig:highz}. Evidently, at redshifts above $z=1$ the density of AGNs in clusters is not dissimilar to the density of AGNs in our sample of lower redshift clusters.  At high-redshifts we probably probe the cores of extended formations, the proto-clusters, which will further collapse to then become  massive clusters. The conditions within the cores of these structures may be similar to the conditions in the cores of our low-mass clusters at much lower redshifts. Therefore, we argue that the AGN fraction in cluster galaxies anti-correlates with cluster mass. 

In brief, our goal was to study the AGN activity in 167 XXL X-ray galaxy clusters as a function of  the mass of the cluster and of  the location of the AGN in the cluster. We found a significant AGN excess in our low-mass cluster subsample between $0.5r_{500,MT}$ and 2$r_{500,MT}$, which decreases to the background level in the cluster cores. In contrast, the high-mass subsample presents no AGN excess, but rather a decreasing trend, consistent with the results of previous studies on massive clusters.

The AGN excess in poor clusters indicates AGN triggering in the outskirts, supporting previous studies that reported enhanced galaxy merging. This effect is probably prevented by high-velocity dispersions in high-mass clusters. 
On the other hand, the AGN density that is consistent with that of the field at the cores of low-mass clusters implies that the AGN suppression mechanism is less effective than the one observed in more massive samples where the AGN density is significantly lower than in the field.

The cluster mass--AGN activity anti-correlation provides evidence of how deeper gravitational potentials prevent AGN triggering in their outskirts and cause more effective ram pressure gas stripping that leads to AGN suppression in their cores.

This study investigates the AGN demographics in galaxy clusters in a statistical way. Further multiwavelength analysis of the individual sources,  as well as a comparison with cosmological hydrodynamical simulations \citep[see][XXL paper XIX]{Koulouridis17}, will be presented
in a companion paper.  

\begin{acknowledgements}
XXL is an international project based around an XMM Very Large Programme surveying two 25 $deg^2$
extragalactic fields at a depth of $5.8\times10^{-15}$ erg s$^{-1}$ cm$^{-2}$ in (0.5--2 keV) at the 90\% completeness level (see XXL paper I).
The XXL website is http://irfu.cea.fr/xxl. Multiband information and spectroscopic follow-up of the
X-ray sources are obtained through a number of survey programmes, summarised at http://xxlmultiwave.pbworks.com/.
We would like to thank A. Mantz and B. Poggianti for comments on the manuscript. The Saclay group acknowledges long-term support from the Centre National d'Etudes Spatiales (CNES). EK thanks CNES and CNRS for support of post-doctoral research. SF acknowledges financial support from the Swiss National Science Foundation. 
This work was supported by the Programme National Cosmology et Galaxies (PNCG) of CNRS/INSU with INP and IN2P3, co-funded by CEA and CNES.
This work is based on observations obtained with MegaPrime/MegaCam, a joint project of CFHT
and CEA/IRFU, at the Canada-France-Hawaii Telescope (CFHT) which is operated by
the National Research Council (NRC) of Canada, the Institut National des Sciences
de l'Univers of the Centre National de la Recherche Scientifique (CNRS) of
France, and the University of Hawaii. This work is based in part on data
products produced at Terapix available at the Canadian Astronomy Data Centre as
part of the Canada-France-Hawaii Telescope Legacy Survey, a collaborative
project of NRC and CNRS. 
\end{acknowledgements}

\bibliographystyle{aa} 
\bibliography{mylib}

\end{document}